\documentclass[sigconf]{acmart}

\AtBeginDocument{%
  }

\expandafter\def\csname ver@subfigure.sty\endcsname{9999/99/99}
\usepackage{subcaption}
\usepackage{multirow}
\usepackage{listings}
\usepackage{colortbl}
\usepackage{pifont}
\usepackage{arydshln}
\usepackage{mathtools}
\usepackage[most]{tcolorbox}
\tcbuselibrary{breakable}
\tcbuselibrary{raster}

\usepackage[capitalize,noabbrev]{cleveref}

\colorlet{punct}{red!60!black}
\definecolor{background}{HTML}{EEEEEE}
\definecolor{delim}{RGB}{20,105,176}
\colorlet{numb}{magenta!60!black}

\lstdefinelanguage{json}{
    basicstyle=\ttfamily\scriptsize,
    numbers=none,
    showstringspaces=false,
    breaklines=true,
    frame=single,
    backgroundcolor=\color{background},
    literate=
     *{0}{{{\color{numb}0}}}{1}
      {1}{{{\color{numb}1}}}{1}
      {2}{{{\color{numb}2}}}{1}
      {3}{{{\color{numb}3}}}{1}
      {4}{{{\color{numb}4}}}{1}
      {5}{{{\color{numb}5}}}{1}
      {6}{{{\color{numb}6}}}{1}
      {7}{{{\color{numb}7}}}{1}
      {8}{{{\color{numb}8}}}{1}
      {9}{{{\color{numb}9}}}{1}
      {:}{{{\color{punct}{:}}}}{1}
      {,}{{{\color{punct}{,}}}}{1}
      {\{}{{{\color{delim}{\{}}}}{1}
      {\}}{{{\color{delim}{\}}}}}{1}
      {[}{{{\color{delim}{[}}}}{1}
      {]}{{{\color{delim}{]}}}}{1},
}

\def\model{CEAE}

\begin{document}

\settopmatter{printfolios=false}
\copyrightyear{2026}
\acmYear{2026}
\setcopyright{cc}
\setcctype{by}
\acmConference[SIGSPATIAL '26]{The 34th ACM International Conference on Advances in Geographic Information Systems}{November 03--06, 2026}{Riverside, CA, USA}
\acmBooktitle{The 34th ACM International Conference on Advances in Geographic Information Systems (SIGSPATIAL '26), November 03--06, 2026, Riverside, CA, USA}
\acmDOI{10.1145/3841645.3843436}
\acmISBN{979-8-4007-2950-8/2026/11}

\title{City Editing: Hierarchical Agentic Execution \\ for Dependency-Aware Urban Geospatial Modification}

\author{Rui Liu}
\affiliation{%
  \institution{University of Kansas}
  \country{}}
\email{rayliu@ku.edu}

\author{Steven Jige Quan}
\affiliation{%
  \institution{Seoul National University}
  \country{}}
\email{sjquan@snu.ac.kr}

\author{Zhong-Ren Peng}
\affiliation{%
  \institution{University of Florida}
  \country{}}
\email{zpeng@dcp.ufl.edu}

\author{Zijun Yao}
\affiliation{%
  \institution{University of Kansas}
  \country{}}
\email{zyao@ku.edu}

\author{Han Wang}
\authornote{Corresponding authors.}
\authornotemark[2]
\affiliation{%
  \institution{University of Kansas}
  \country{}}
\email{han.wang@ku.edu}

\author{Pengyang Wang}
\affiliation{%
  \institution{University of Macau}
  \country{}}
\email{pywang@um.edu.mo}

\author{Kunpeng Liu}
\affiliation{%
  \institution{Clemson University}
  \country{}}
\email{kunpenl@clemson.edu}

\author{Yanjie Fu}
\affiliation{%
  \institution{Arizona State University}
  \country{}}
\email{yanjie.fu@asu.edu}

\author{Dongjie Wang}
\authornotemark[2]
\affiliation{%
  \institution{University of Kansas}
  \country{}}
\email{wangdongjie@ku.edu}


\renewcommand{\shortauthors}{Rui Liu et al.}

\begin{abstract}
Urban renewal requires incremental modifications to existing geospatial plans, yet manually updating complex layouts under spatial constraints is labor-intensive and error-prone.
To tackle this, we propose \model, a hierarchical agentic framework that formulates urban renewal as machine-executable GeoJSON editing from natural-language instructions.
\model~decomposes instructions into hierarchical geometric intents, executing edits coarse-to-fine while preserving spatial consistency via a self-reflective execution–validation loop.
Experimental results show that {\model} outperforms baselines in execution validity, robustness, and geometric accuracy.
\end{abstract}

\maketitle

\vspace{-0.3cm}
\section{Introduction}
As urban environments evolve, cities increasingly exhibit localized deficiencies such as traffic congestion and functional imbalance. Addressing these issues rarely involves redesigning urban environments from scratch; instead, contemporary urban renewal marks a paradigm shift from holistic plan generation toward incremental refinement of existing layouts. However, urban renewal in practice remains highly deliberative and manual. Even minor updates, such as expanding a park boundary, require planners to manually adjust nearby paths and relocate street furniture to satisfy geometric validity and spatial buffer constraints. Consequently, current manual workflows are labor-intensive, error-prone, and difficult to scale. Despite recent advances in urban intelligence, existing computational paradigms, including generative models~\cite{liu2025urbanplaningaiagent}, optimization-based methods~\cite{wang2023hierarchical}, and standard LLM agents~\cite{zhou2024largelanguagemodelparticipatory}, primarily treat urban planning as a one-shot holistic generation task. They fail to support the iterative, dependency-aware modifications required for real-world urban renewal, as they lack explicit mechanisms to model decision dependencies and propagate spatial constraints.

To bridge this gap, we formalize urban renewal as an automated city editing process operating directly on existing GeoJSON layouts. Realizing such a framework requires addressing three core challenges: 1) defining machine-executable editing operations over structured geospatial data, 2) propagating spatial constraints across editing stages to handle inter-entity dependencies, and 3) systematically detecting and mitigating intermediate geometric errors to prevent cascading failures. To tackle these challenges, we propose \textbf{\model}, a hierarchical agentic execution framework. \model~adopts a coarse-to-fine execution strategy across polygon-, line-, and point-level representations, ensuring that fine-grained edits inherit validated spatial constraints from coarser levels. To guarantee execution reliability, \model~integrates a self-reflective validator after each subtask, utilizing diagnostic feedback to trigger corrective re-execution and prevent error propagation. In summary, our main contributions are threefold: 1) formalizing city geospatial editing as a structured computational paradigm for dependency-aware layout modification, 2) proposing a hierarchical multi-agent framework with coarse-to-fine constraint propagation and self-reflective validation, and 3) providing comprehensive empirical evidence that the proposed framework yields consistent gains in editing correctness, structural validity, and robustness across diverse urban scenarios.

\begin{figure*}[!ht]
    \centering
    \includegraphics[width=\linewidth]{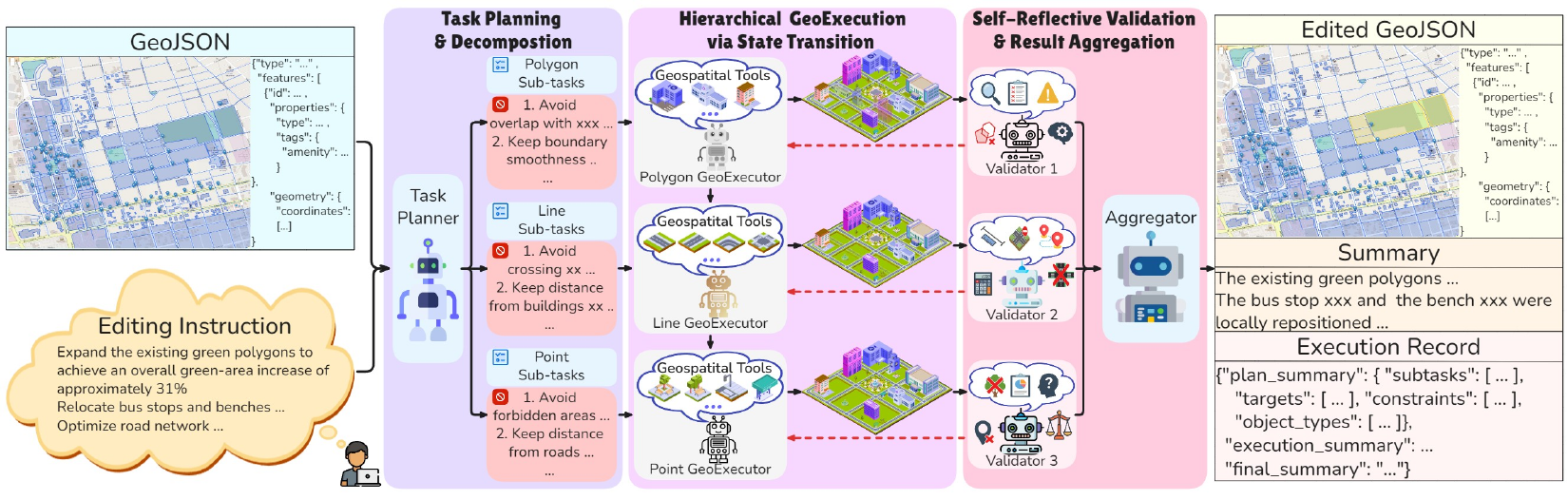}
    \vspace{-0.7cm}
    \caption{Overview of \model. Given a GeoJSON layout and a natural-language instruction, \model~decomposes editing intents and executes them coarse-to-fine with iterative validation and constraint propagation.}


    \label{fig:framework}
\end{figure*}

\section{Preliminaries \& Problem Statement}
We represent an urban layout as a GeoJSON object $G$ containing geometric elements (polygons, lines, points) and spatial attributes, and let $P$ denote a natural-language editing instruction specifying desired layout modifications.
The goal of city geospatial editing is to realize a mapping $f: (G, P) \rightarrow G^*$ that updates the layout while preserving geometric and topological consistency.
Rather than executing edits in a single pass, the system follows a coarse-to-fine execution sequence across geometric abstraction levels:
\vspace{-0.2cm}
\begin{equation}
    G^{(0)} \xrightarrow{\;I_{\mathrm{poly}}\;}
    G^{(1)} \xrightarrow{\;I_{\mathrm{line}}\;}
    G^{(2)} \xrightarrow{\;I_{\mathrm{point}}\;}
    G^*,
    \vspace{-0.3cm}
\end{equation}
where $G^{(0)} = G$ is the initial layout, and $I_{\ell}$ represents the set of editing intents at geometric level $\ell \in \{\mathrm{polygon}, \mathrm{line}, \mathrm{point}\}$.
Edits at coarser geometric levels produce updated layouts that condition subsequent finer-grained edits, explicitly propagating structural constraints to produce a valid output $G^*$.

\vspace{-0.2cm}
\section{Methodology}
\vspace{-0.1cm}
\subsection{Framework Overview}
Figure~\ref{fig:framework} shows the overall architecture of {\model}, which coordinates four specialized agent roles: a task planner, geoexecutors, validators, and an aggregator.
Given a GeoJSON layout and a natural-language instruction, {\model} decomposes high-level requests into hierarchical geometric intents, executing them coarse-to-fine via level-specific geospatial tools while propagating intermediate spatial constraints.
Throughout execution, self-reflective validation verifies intermediate states to prevent error accumulation, after which the aggregator consolidates all validated edits into the final updated GeoJSON plan.

\vspace{-0.3cm}
\subsection{Task Planning and Editing Intent Decomposition}
High-level natural-language instructions often involve implicit spatial constraints and multi-entity operations.
Direct execution without explicit decomposition easily leads to constraint violations and invalid intermediate states.
To address this, the task planner $\mathcal{F}_{\mathrm{plan}}$ interprets the instruction $P$ over layout $G$ to induce a structured editing plan $\Pi = \mathcal{F}_{\mathrm{plan}}(P, G) = \{ I_{\ell} \mid \ell \in \{\mathrm{polygon}, \mathrm{line}, \mathrm{point}\} \}$.
Here, each geometry-level editing intent $I_{\ell}$ serves as a composite planning unit, defined as:
\vspace{0.1cm}
\begin{equation}
  I_{\ell} =
  \left(
  \mathrm{scope}_{\ell},\;
  \mathrm{goal}_{\ell},\;
  \mathcal{T}_{\ell},\;
  \mathcal{C}_{\ell}
  \right),
\end{equation}
where $\mathrm{scope}_{\ell}$ specifies target spatial entities, $\mathrm{goal}_{\ell}$ defines editing objectives, $\mathcal{T}_{\ell} = \{\tau_{\ell,k}\}_{k=1}^{K_{\ell}}$ denotes the ordered subtask sequence, and $\mathcal{C}_{\ell} = \{c_{\ell,j}\}_{j=1}^{J}$ represents required intent-level constraints.
This structured formulation explicitly connects high-level semantic intent with geometry-level execution.

\subsection{Hierarchical GeoExecution via State Transition}
To handle complex spatial interdependencies, {\model} executes planned subtasks through a coarse-to-fine state transition mechanism across polygon-, line-, and point-level representations.
At each geometric level $\ell \in \{\mathrm{polygon}, \mathrm{line}, \mathrm{point}\}$, given the subtask sequence $\mathcal{T}_{\ell} = \{\tau_{\ell,1}, \dots, \tau_{\ell,K_{\ell}}\}$, execution initializes at state $G_{\ell,1} = G^{(\ell)}$.
For each subtask $\tau_{\ell,k}$, the geoexecutor invokes deterministic geospatial operators to yield an observation--state pair:
\begin{equation}
  \label{eq:geoexec-transition}
  (O_{\ell,k}, G_{\ell,k+1}) = \mathcal{F}_{\mathrm{exec}}^{\ell}(\tau_{\ell,k}, G_{\ell,k}),
  \;\; {\scriptstyle k = 1, \dots, K_{\ell}}
\end{equation}
where each subtask produces either an informational observation or a state update, but not both: $O_{\ell,k}$ records execution observations for informational subtasks (where $G_{\ell,k+1}=G_{\ell,k}$), while it is empty for state-updating subtasks that modify the geometry $G_{\ell,k+1}$.
Upon completing all subtasks at level $\ell$, the updated layout $G^{(\ell+1)} = G_{\ell,K_{\ell}+1}$ is propagated as the input context to the next finer level. 

\begin{table*}[!t]
    \centering
    \caption{Main comparison results on point, line, and polygon editing tasks. Single-pass denotes one-shot execution, in which the LLM directly produces a full sequence of editing operations without an explicit task planner or a validator. \textbf{Bold} indicates the best performance, \underline{underline} indicates the second best, and all results are reported as mean$\pm$standard deviation.}
    \vspace{-0.2cm}
    \label{tab:main_results}
    \resizebox{\textwidth}{!}{
        \begin{tabular}{llccccccc}
            \toprule
            \multirow{2}{*}{{Editing Paradigm}}
                                                  & \multirow{2}{*}{Backbone}
                                                  & \multicolumn{2}{c}{Point Editing}
                                                  & \multicolumn{2}{c}{Line Editing}
                                                  & \multicolumn{2}{c}{Polygon Editing}                                                                                                                                                                                                       \\
            \cmidrule(lr){3-4} \cmidrule(lr){5-6} \cmidrule(lr){7-8}
                                                  &
                                                  & \textbf{L1-REE $\downarrow$}        & \textbf{L1-EVR $\uparrow$}     & \textbf{L2-REE $\downarrow$}   & \textbf{L2-EVR $\uparrow$}     & \textbf{L3-ACE $\downarrow$}   & \textbf{L3-EVR $\uparrow$}                                      \\
            \midrule
            \multicolumn{8}{c}{\cellcolor{gray!25}\textit{Closed-source LLMs}}                                                                                                                                                                                                                \\
            \noalign{\vskip 0.3em}
            Single-pass                           & GPT-5-Mini                          & 0.234\tiny{±0.476}             & \textbf{0.984\tiny{±0.007}}    & 0.502\tiny{±0.192}             & \underline{0.945\tiny{±0.023}} & \underline{0.097\tiny{±0.047}} & 0.951\tiny{±0.013}             \\
            Single-pass                           & GPT-5-Nano                          & 0.265\tiny{±0.395}             & 0.942\tiny{±0.009}             & 0.545\tiny{±0.283}             & 0.936\tiny{±0.032}             & 0.108\tiny{±0.136}             & 0.902\tiny{±0.029}             \\
            \midrule
            \multicolumn{8}{c}{\cellcolor{gray!25}\textit{Open-source LLMs}}                                                                                                                                                                                                                  \\
            \noalign{\vskip 0.3em}

            Single-pass                           & DeepSeek-R1-Distill-Qwen-1.5B       & \underline{0.197\tiny{±0.206}} & 0.716\tiny{±0.083}             & \underline{0.485\tiny{±0.226}} & 0.658\tiny{±0.121}             & 0.391\tiny{±0.425}             & 0.713\tiny{±0.127}             \\
            Single-pass                           & DeepSeek-R1-Distill-Qwen-7B         & 0.549\tiny{±0.351}             & 0.846\tiny{±0.017}             & 0.763\tiny{±0.176}             & 0.925\tiny{±0.015}             & 0.241\tiny{±0.392}             & 0.844\tiny{±0.072}             \\
            Single-pass                           & DeepSeek-R1-Distill-Qwen-14B        & 0.306\tiny{±0.442}             & 0.926\tiny{±0.037}             & 0.594\tiny{±0.231}             & 0.893\tiny{±0.039}             & 0.135\tiny{±0.233}             & 0.859\tiny{±0.115}             \\
            Single-pass                           & DeepSeek-R1-Distill-Llama-8B        & 0.286\tiny{±0.489}             & 0.866\tiny{±0.024}             & 0.537\tiny{±0.427}             & 0.852\tiny{±0.059}             & 0.159\tiny{±0.324}             & 0.926\tiny{±0.048}             \\
            \noalign{\vskip 0.3em}
            \hdashline
            \noalign{\vskip 0.3em}
            Single-pass                           & Qwen2.5-3B-Instruct                 & 0.839\tiny{±0.254}             & 0.932\tiny{±0.018}             & 0.862\tiny{±0.074}             & 0.878\tiny{±0.016}             & 0.190\tiny{±0.304}             & 0.787\tiny{±0.107}             \\
            Single-pass                           & Qwen2.5-7B-Instruct                 & 0.794\tiny{±0.186}             & 0.947\tiny{±0.024}             & 0.803\tiny{±0.180}             & 0.885\tiny{±0.027}             & 0.143\tiny{±0.268}             & \underline{0.971\tiny{±0.011}} \\
            Single-pass                           & Qwen3-8B                            & 0.201\tiny{±0.430}             & 0.702\tiny{±0.124}             & 0.663\tiny{±0.259}             & 0.592\tiny{±0.060}             & 0.125\tiny{±0.214}             & 0.659\tiny{±0.093}             \\
            \noalign{\vskip 0.3em}
            \hdashline
            \noalign{\vskip 0.3em}
            Single-pass                           & Llama-3.1-8B-Instruct               & 0.214\tiny{±0.452}             & 0.951\tiny{±0.019}             & 0.492\tiny{±0.542}             & 0.941\tiny{±0.011}             & 0.126\tiny{±0.263}             & 0.924\tiny{±0.029}             \\
            Single-pass                           & LLaMA-3-8B-Instruct                 & 0.470\tiny{±0.746}             & 0.938\tiny{±0.014}             & 0.562\tiny{±0.461}             & 0.906\tiny{±0.031}             & 0.145\tiny{±0.210}             & 0.706\tiny{±0.196}             \\
            \rowcolor{pink!50} Agentic ({\model}) & LLaMA-3-8B-Instruct                 & \textbf{0.187\tiny{±0.235}}    & \underline{0.975\tiny{±0.019}} & \textbf{0.319\tiny{±0.396}}    & \textbf{0.947\tiny{±0.042}}    & \textbf{0.081\tiny{±0.105}}    & \textbf{0.979\tiny{±0.017}}    \\
            \bottomrule
        \end{tabular}
    }
\end{table*}
\subsection{Self-Reflective Validation and Result Aggregation}
To prevent intermediate geometric errors from propagating across execution stages, each geoexecutor is paired with a validator $\mathcal{F}_{\mathrm{val}}$ that acts as an admissibility gate.
After subtask $\tau_{\ell,k}$, the validator evaluates the outcome $(O_{\ell,k}, G_{\ell,k+1})$ under state $G_{\ell,k}$ to output a binary decision:
\begin{equation}
  \label{eq:val-decision}
  V_{\ell,k} = \mathcal{F}_{\mathrm{val}}\!\left(
  I_{\ell},\;
  \tau_{\ell,k},\;
  G_{\ell,k},\;
  O_{\ell,k},\;
  G_{\ell,k+1}
  \right) \in \{\text{accept}, \text{reject}\}.
\end{equation}
Validation criteria rely on deterministic operators based on Shapely~\cite{Gillies_Shapely_2025} and GeoPandas~\cite{kelsey_jordahl_2020_3946761} to verify geometric validity, intent compliance, and boundary constraints.
If $V_{\ell,k} = \text{accept}$, the updated state is committed. 
Otherwise, structured diagnostic feedback is generated to trigger subtask re-execution with maximum retries $R$.
After completing all stages, the aggregator $\mathcal{F}_{\mathrm{agg}}$ integrates validated outcomes to synthesize the final plan $G^*$ and execution summary $S$:
\vspace{-0.2cm}
\begin{equation}
    \vspace{-0.2cm}
  \label{eq:aggregator}
  (G^{*}, S) =
  \mathcal{F}_{\mathrm{agg}}\!\left(
  \{((O_{\ell,k}, G_{\ell,k+1}), V_{\ell,k})\}
  \right).
\end{equation}
Here, $G^*$ commits only validated state-updating modifications, while $S$ consolidates subtask-level traces, validation decisions, and retry statistics to ensure full execution traceability.



\begin{figure}[!t]
    \vspace{-0.1cm}
    \centering
    \includegraphics[width=\linewidth]{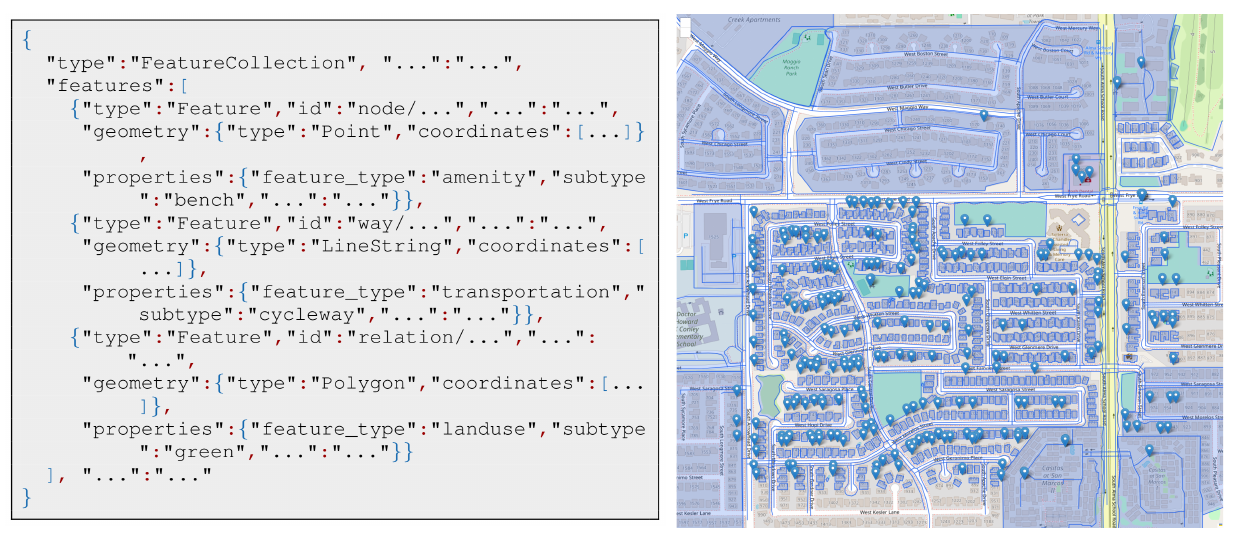}
    \vspace{-0.7cm}
    \caption{An illustration of GeoJSON (left) and the corresponding geometric visualization (right).}
    \label{fig:appendix_geojson_example}
    \vspace{-0.3cm}
\end{figure}
\vspace{-0.3cm}
\section{Experiment}

\subsection{Experimental Setup}
\subsubsection{Datasets and Baselines.}
\label{sec:dataset_construction}
\label{sec:baseline_model}
We construct urban geospatial editing benchmarks by extracting 2,500 urban patches (1~km $\times$ 1~km scale) across 219 cities from OpenStreetMap (OSM)~\cite{openstreetmap}.
The dataset comprises 500 Level-1 point editing, 500 Level-2 line editing, and 1,500 Level-3 polygon editing samples, with a larger polygon allocation to ensure adequate coverage of the greater geometric and constraint variability at this level.
Ground-truth editing targets are generated via deterministic spatial transformations, while natural-language instructions are produced semi-automatically from manually crafted seed templates, followed by LLM-assisted paraphrasing and expansion, and human inspection.
To assess editing performance, we compare \model~against a broad set of single-pass models spanning diverse families and parameter counts, including closed-source models (GPT-5-Mini, GPT-5-Nano) and open-source backbones spanning 1.5B to 14B parameters (LLaMA-3/3.1~\cite{llama3modelcard}, Qwen2.5/3~\cite{qwen2.5}, and DeepSeek-R1 distillations~\cite{deepseekai2025deepseekr1incentivizingreasoningcapability}). 

\vspace{-0.2cm}
\subsubsection{Evaluation Metrics and Implementation.}
\label{sec:experimental_settings}
We evaluate performance across task levels using three complementary metrics.
For point and line editing, \emph{Relative Execution Error (REE)} measures geometric deviation relative to target edit magnitude $m$:
\vspace{-0.1cm}
\begin{equation}
    \mathrm{REE} = \frac{d(g_{\mathrm{edit}}, g_{\mathrm{label}})}{m},
\end{equation}

where $d(\cdot)$ computes local planar Euclidean distance in meters.
For polygon editing, \emph{Area Compliance Error (ACE)} assesses target area change ratio compliance:
\vspace{-0.1cm}
\begin{equation}
    \mathrm{ACE} = \frac{|\Delta A_{\mathrm{green}} - A_{\mathrm{target}}|}{A_{\mathrm{target}}}.
\end{equation}
To quantify end-to-end execution reliability, \emph{Execution Validity Rate (EVR)} measures the fraction of task executions that complete without runtime errors or geometric violations:
\vspace{-0.1cm}
\begin{equation}
    \mathrm{EVR} = \frac{1}{N} \sum_{n=1}^{N} \mathbb{I}\!\left( \mathcal{E}^{(n)} = \text{valid} \right).
    \vspace{-0.1cm}
\end{equation}
All experiments were conducted on an AMD Ryzen Threadripper PRO 7965WX CPU and an NVIDIA RTX 6000 Ada GPU.
We adopt deterministic decoding ($\texttt{do\_sample}=\texttt{False}$) for Level-1/2 tasks and temperature sampling ($\texttt{temperature}=0.9, \texttt{top\_p}=0.9$) for Level-3 planning, with subtask validation retries bounded by $R=3$.
\label{sec:hyperparameters}

\vspace{-0.1cm}
\subsection{Experimental Results}
\subsubsection{Overall Performance.}
Table~\ref{tab:main_results} compares {\model} with single-pass baselines across three task levels.
{\model} consistently achieves lower geometric errors and higher execution validity rates (EVR), with performance gains widening on more complex line and polygon tasks.
This superiority demonstrates the joint advantage of hierarchical intent planning, coarse-to-fine constraint propagation, and reflective validation.
\vspace{-0.2cm}
\subsubsection{Ablation Study.}
\begin{figure}[!t]
    \centering
    \includegraphics[width=\linewidth]{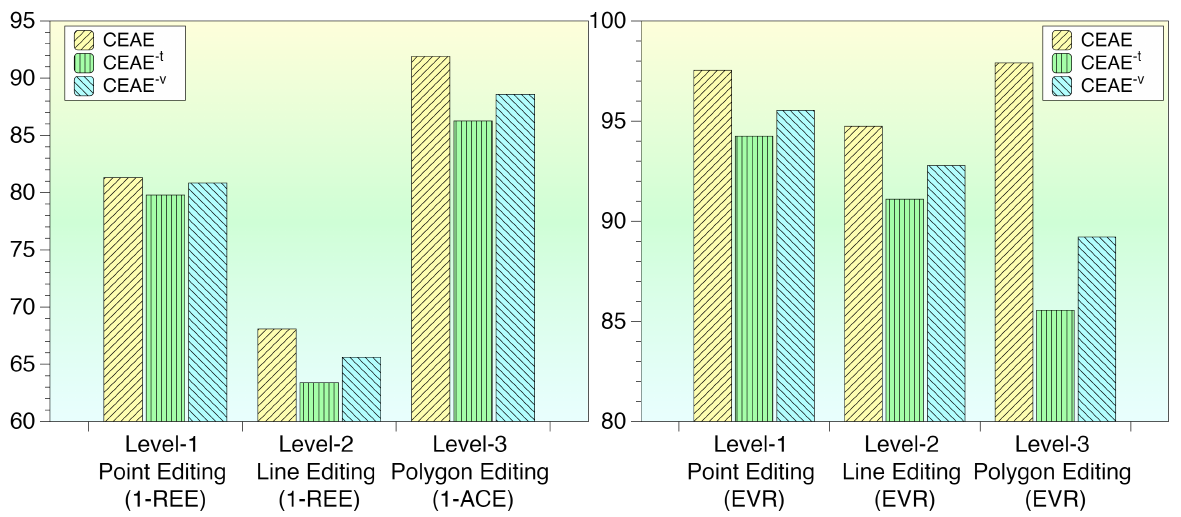}
    \vspace{-0.65cm}
    \caption{Ablation study evaluating the removal of the task planner ($-t$) and validator ($-v$) across three task levels.}
    \label{fig:ablation_study}
    \vspace{-0.3cm}
\end{figure}

We evaluate variants omitting the task planner ({\model}$^{-t}$, where the raw instruction is directly forwarded to geoexecutors without decomposition) or validator ({\model}$^{-v}$).
As Fig.~\ref{fig:ablation_study} shows, {\model} consistently outperforms both variants across all levels, with the gap expanding as task complexity grows.
These results confirm that explicit planning aligns execution with global intents, while reflective validation prevents error propagation.

\section{Related Work}
Urban planning has attracted increasing attention in computational research~\cite{wang2025generativeaimeetsfuture,liu2025urbanplaningaiagent}.
\textbf{Urban Spatial Planning Models.}
Urban planning has been widely studied from a generative perspective, where models learn spatial patterns to synthesize land-use configurations or block layouts~\cite{wang2020reimagining,wang2023automated,wang2021deep,wang2024dual}.
Further studies incorporated hierarchical reinforcement learning or deep generative frameworks to model functional dependencies across regional and block levels~\cite{wang2023human,wang2023hierarchical}.
However, these methods focus primarily on generating complete layouts from scratch or holistic optimization.
They lack mechanisms for instruction-driven incremental editing over existing geospatial plans, and do not explicitly propagate and validate intermediate editing states across geometric abstraction levels.
\textbf{LLM Agents and Skill Abstractions.}
Large language models (LLMs) have demonstrated strong reasoning capabilities, serving as decision-making interfaces for complex tasks~\cite{10.1145/3820356.3820363}.
Recent agentic AI architectures integrate LLMs into structured workflows combining planning, tool execution, and reflective feedback~\cite{liu2026agentosapplicationsilosnatural}.
Moreover, recent work extends agent tool-use into composable, reusable skill abstractions to formalize domain-specific capabilities~\cite{xu2026agentskills,zheng2025skillweaver}.
Building on these advances, {\model} integrates natural-language intent parsing with deterministic geospatial tools, enabling reliable, dependency-aware city editing through structured skill-level execution.

\section{Conclusion}
In this work, we formalize city geospatial editing and propose {\model}, a hierarchical agentic framework that transforms natural-language renewal requests into structured GeoJSON edits via coarse-to-fine execution and self-reflective validation.
Extensive experiments show that structured intent decomposition and execution-level validation are essential for preventing error propagation across complex spatial tasks.
Future work will explore incorporating richer domain constraints and interactive human feedback to enhance adaptability across diverse urban renewal scenarios.

\bibliographystyle{ACM-Reference-Format}
\bibliography{main}

@inproceedings{wang2020reimagining,
  title     = {Reimagining city configuration: Automated urban planning via adversarial learning},
  author    = {Wang, Dongjie and Fu, Yanjie and Wang, Pengyang and Huang, Bo and Lu, Chang-Tien},
  booktitle = {Proceedings of the 28th international conference on advances in geographic information systems},
  pages     = {497--506},
  year      = {2020}
}

@misc{wang2025generativeaimeetsfuture,
  title         = {Generative AI Meets Future Cities: Towards an Era of Autonomous Urban Intelligence},
  author        = {Dongjie Wang and Chang-Tien Lu and Xinyue Ye and Tan Yigitcanlar and Yanjie Fu},
  year          = {2025},
  eprint        = {2304.03892},
  archiveprefix = {arXiv},
  primaryclass  = {cs.AI},
  url           = {https://arxiv.org/abs/2304.03892}
}

@article{wang2023automated,
  title     = {Automated urban planning for reimagining city configuration via adversarial learning: quantification, generation, and evaluation},
  author    = {Wang, Dongjie and Fu, Yanjie and Liu, Kunpeng and Chen, Fanglan and Wang, Pengyang and Lu, Chang-Tien},
  journal   = {ACM Transactions on Spatial Algorithms and Systems},
  volume    = {9},
  number    = {1},
  pages     = {1--24},
  year      = {2023},
  publisher = {ACM New York, NY}
}

@inproceedings{wang2021deep,
  title        = {Deep human-guided conditional variational generative modeling for automated urban planning},
  author       = {Wang, Dongjie and Liu, Kunpeng and Johnson, Pauline and Sun, Leilei and Du, Bowen and Fu, Yanjie},
  booktitle    = {2021 IEEE international conference on data mining (ICDM)},
  pages        = {679--688},
  year         = {2021},
  organization = {IEEE}
}

@inproceedings{wang2023human,
  title     = {Human-instructed deep hierarchical generative learning for automated urban planning},
  author    = {Wang, Dongjie and Wu, Lingfei and Zhang, Denghui and Zhou, Jingbo and Sun, Leilei and Fu, Yanjie},
  booktitle = {Proceedings of the AAAI Conference on Artificial Intelligence},
  volume    = {37},
  number    = {4},
  pages     = {4660--4667},
  year      = {2023}
}

@inproceedings{wang2023hierarchical,
  title        = {Hierarchical reinforced urban planning: Jointly steering region and block configurations},
  author       = {Wang, Pengfei and Wang, Daniel and Liu, Kunpeng and Wang, Dongjie and Zhou, Yuanchun and Sun, Leilei and Fu, Yanjie},
  booktitle    = {Proceedings of the 2023 SIAM International Conference on Data Mining (SDM)},
  pages        = {343--351},
  year         = {2023},
  organization = {SIAM}
}

@inproceedings{wang2024dual,
  author    = {Xuanming Hu and Wei Fan and Dongjie Wang and Pengyang Wang and Yong Li and Yanjie Fu},
  title     = {Dual-stage Flows-based Generative Modeling for Traceable Urban Planning},
  booktitle = {Proceedings of the 2024 SIAM International Conference on Data Mining (SDM)},
  chapter   = {},
  pages     = {370-378},
  year      = {2024},
  doi       = {10.1137/1.9781611978032.42}
}

@misc{openstreetmap,
  author       = {{OpenStreetMap}},
  title        = {{OpenStreetMap}},
  howpublished = {\url{https://www.openstreetmap.org}},
  year         = {2025}
}

@article{llama3modelcard,
  title  = {Llama 3 Model Card},
  author = {AI@Meta},
  year   = {2024},
  url    = {https://github.com/meta-llama/llama3/blob/main/MODEL_CARD.md}
}

@misc{qwen2.5,
  title  = {Qwen2.5: A Party of Foundation Models},
  url    = {https://qwenlm.github.io/blog/qwen2.5/},
  author = {Qwen Team},
  month  = {September},
  year   = {2024}
}

@misc{deepseekai2025deepseekr1incentivizingreasoningcapability,
  title         = {DeepSeek-R1: Incentivizing Reasoning Capability in LLMs via Reinforcement Learning},
  author        = {DeepSeek-AI},
  year          = {2025},
  eprint        = {2501.12948},
  archiveprefix = {arXiv},
  primaryclass  = {cs.CL},
  url           = {https://arxiv.org/abs/2501.12948}
}

@article{liu2025urbanplaningaiagent,
  author     = {Liu, Rui and Zhe, Tao and Peng, Zhong-Ren and Catbas, Necati and Ye, Xinyue and Wang, Dongjie and Fu, Yanjie},
  title      = {Urban Planning in the Age of Agentic AI: Emerging Paradigms and Prospects},
  year       = {2025},
  issue_date = {December 2025},
  publisher  = {Association for Computing Machinery},
  address    = {New York, NY, USA},
  volume     = {27},
  number     = {2},
  issn       = {1931-0145},
  url        = {https://doi.org/10.1145/3787470.3787474},
  doi        = {10.1145/3787470.3787474},
  journal    = {SIGKDD Explor. Newsl.},
  month      = dec,
  pages      = {35–42},
  numpages   = {8}
}

@misc{zhou2024largelanguagemodelparticipatory,
  title         = {Large Language Model for Participatory Urban Planning},
  author        = {Zhilun Zhou and Yuming Lin and Depeng Jin and Yong Li},
  year          = {2024},
  eprint        = {2402.17161},
  archiveprefix = {arXiv},
  primaryclass  = {cs.AI},
  url           = {https://arxiv.org/abs/2402.17161}
}

@misc{xu2026agentskills,
  title         = {Agent Skills for Large Language Models: Architecture, Acquisition, Security, and the Path Forward},
  author        = {Xu, Renjun and Yan, Yang},
  year          = {2026},
  eprint        = {2602.12430},
  archiveprefix = {arXiv},
  primaryclass  = {cs.MA},
  url           = {https://arxiv.org/abs/2602.12430}
}

@misc{zheng2025skillweaver,
  title         = {SkillWeaver: Web Agents can Self-Improve by Discovering and Honing Skills},
  author        = {Zheng, Boyuan and Fatemi, Michael Y. and Jin, Xiaolong and Wang, Zora Zhiruo and Gandhi, Apurva and Song, Yueqi and Gu, Yu and Srinivasa, Jayanth and Liu, Gaowen and Neubig, Graham and Su, Yu},
  year          = {2025},
  eprint        = {2504.07079},
  archiveprefix = {arXiv},
  primaryclass  = {cs.AI},
  url           = {https://arxiv.org/abs/2504.07079}
}

@software{Gillies_Shapely_2025,
  author  = {Gillies, Sean and van der Wel, Casper and Van den Bossche, Joris and Taves, Mike W. and Arnott, Joshua and Ward, Brendan C. and {others}},
  doi     = {10.5281/zenodo.5597138},
  license = {BSD-3-Clause},
  month   = may,
  title   = {{Shapely}},
  url     = {https://github.com/shapely/shapely},
  version = {2.1.1},
  year    = {2025}
}

@software{kelsey_jordahl_2020_3946761,
  author    = {Jordahl, Kelsey and Van den Bossche, Joris and Fleischmann, Martin and others},
  title     = {geopandas/geopandas: v0.8.1},
  year      = {2020},
  month     = jul,
  publisher = {Zenodo},
  version   = {v0.8.1},
  doi       = {10.5281/zenodo.3946761},
  url       = {https://doi.org/10.5281/zenodo.3946761}
}

@article{10.1145/3820356.3820363,
author = {Liu, Kunpeng and Nakata, Nori and Zhang, Jinghan and Chen, Guodong and Liu, Rui and Zhe, Tao and Wang, Dongjie and Wang, Xinyuan and Cao, Hongyu and Fu, Yanjie},
title = {Beyond Simulate-Then-Optimize: Geothermal AI for Geothermal Dynamics Prediction, Design, and Discovery},
year = {2026},
issue_date = {June 2026},
publisher = {Association for Computing Machinery},
address = {New York, NY, USA},
volume = {28},
number = {1},
issn = {1931-0145},
url = {https://doi.org/10.1145/3820356.3820363},
doi = {10.1145/3820356.3820363},
journal = {SIGKDD Explor. Newsl.},
month = jun,
pages = {102–114},
numpages = {13}
}

@misc{liu2026agentosapplicationsilosnatural,
      title={AgentOS: From Application Silos to a Natural Language-Driven Data Ecosystem}, 
      author={Rui Liu and Tao Zhe and Dongjie Wang and Zijun Yao and Kunpeng Liu and Yanjie Fu and Huan Liu and Jian Pei},
      year={2026},
      eprint={2603.08938},
      archivePrefix={arXiv},
      primaryClass={cs.AI},
      url={https://arxiv.org/abs/2603.08938}, 
}

\newpage
\appendix

\end{document}